\documentclass[prb,preprint]{revtex4}

\usepackage{amsmath}

\usepackage{graphicx}

\newcommand{\ve}[1]{{\overrightarrow{#1}}}
\newcommand{\cbare}[1]{{\overline{\mathcal{#1}}}}
\newcommand{\bare}[1]{{\overline{#1}}}
\begin{document}

\title{Generalized composition law from $2\times2$ matrices}

\author{R. Giust}
\email{remo.giust@univ-fcomte}
\affiliation{Institut FEMTO-ST, UMR CNRS 6174, Universit\'e de Franche-Comt\'e, 16 Route de Gray, 25030 Besan\c con
Cedex France}

\author{J.-M. Vigoureux}
\email{jean-marie.vigoureux@univ-fcomte.fr}
\affiliation{Institut UTINAM, UMR CNRS 6213, Universit\'e de
Franche-Comt\'e, 16 Route de Gray, 25030 Besan\c{c}on Cedex France}

\author{J. Lages}\email{jose.lages@utinam.cnrs.fr}
\affiliation{Institut UTINAM, UMR CNRS 6213, Universit\'e de
Franche-Comt\'e, 16 Route de Gray, 25030 Besan\c{c}on Cedex France}

\begin{abstract}
Many results that are difficult can be found more easily by using a generalization in the complex plane of Einstein's
addition law of parallel velocities. Such a generalization is a natural way to add quantities that are limited to
bounded values. We show how this generalization directly provides phase factors such as the Wigner angle in special
relativity and how this generalization is connected in the simplest case with the composition of $2\times2$ $S$
matrices.
\end{abstract}

\maketitle

\section{Introduction}
In special relativity the composition law of parallel velocities appears to be the natural addition law for quantities whose
values are limited to the closed interval $[-1,1]$, where we have set the speed of light $c = 1$. It
is natural to generalize the composition law of parallel velocities to the complex plane as
\begin{equation}\label{complaw}
A=A_2\oplus A_1=\frac{A_1+A_2}{1+\overline{A_2}\,A_1},
\end{equation}
where $A_1$ and $A_2$ are complex quantities and where the denominator appears as a normalization term (if not otherwise stated, $\overline{A}$ denotes the complex conjugation operation). The physical meaning of this composition law is similar to that of the composition law of parallel velocities in special relativity. Equation~\eqref{complaw} shows that no
matter what real values we give to $A_1 = v_1$ and $A_2 = v_2$, subject only to $v_1<c$ and $v_2<c$, the value
of the resulting velocity $A = w$ cannot exceed the speed of light $c=1$. In the same way, no matter the values of
the complex quantities $A_1$ and $A_2$ (subject only to $|A_1|<1$ and $|A_2|<1$), the
modulus of the resulting quantity $A$ cannot exceed unity.

Because it avoids infinities, such a generalization of Einstein's composition law of velocities appears to be a natural
addition law in a closed interval. As expected, it reduces to the usual addition of
arithmetics when the quantities are small. As shown in Refs.~\onlinecite{Vigoureux01,Vigoureux92,Lages08}, the use of this composition law quickly leads to important theoretical results
and provides useful algorithms for computer calculations. The use of Eq.~\eqref{complaw} also leads to results which converge more rapidly than by using transfer
matrices.\cite{Grossel94}

\section{Some simple examples}\label{somex}

We consider three examples\cite{Vigoureux01,Vigoureux92,Lages08} where the use of Eq.~\eqref{complaw} is useful. The examples are
the composition of two non-parallel velocities in special relativity, the reflection
coefficient of a Fabry-Perot in optics, and the characteristics of a polarizer resulting from the association of two successive non-perfect polarizers.

The composition law for two parallel velocities
$v_1$ and $v_2$ in special relativity is ($c=1$)
\begin{equation}\label{compvitesse}
w=v_2\oplus v_1=\frac{v_1+v_2}{1+v_1\,v_2}.
\end{equation}
The calculation of the resulting velocity of two parallel velocities is straightforward. However, it is not when the two velocities are not parallel for which calculations may be tedious. They
become simple when we consider Eq.~\eqref{complaw} which is the generalization in the complex plane of Eq.~\eqref{compvitesse}.
As explained in Ref.~\onlinecite{Vigoureux01}, we replace each velocity $\ve{v_i}$ by the complex number
\begin{equation}\label{defvitesse}
V_i=\tanh\displaystyle\frac{a_i}{2}\,e^{i\alpha_i},
\end{equation}
where the rapidity $a_i$ is related to the modulus of $\ve{v_i}$ by $\tanh{a_i}= v_i$ and where the phase $\alpha_i$
gives the orientation of $\ve{v_i}$ with respect to an arbitrary axis of the reference frame of the observer in the
plane of $\ve{v_1}$ and $\ve{v_2}$. The modulus and phase $\alpha$ of the velocity
$\ve{w}$ resulting from the relativistic composition of $\ve{v_1}$ and $\ve{v_2}$ is directly obtained
\cite{Vigoureux01} by using Eq.~\eqref{complaw}
\begin{equation}\label{non parallel vel}
W=\tanh\displaystyle\frac{a}{2}\,e^{i\alpha}= V_2 \oplus V_1 = \frac{\tanh\displaystyle\frac{a_1}{2}\,e^{i\alpha_1}+ \tanh\displaystyle\frac{a_2}{2}\,e^{i\alpha_2}}{1+\tanh\displaystyle\frac{a_2}{2}\,e^{-i\alpha_2}\, \tanh\displaystyle\frac{a_1}{2}\,e^{i\alpha_1}}.
\end{equation}
The modulus and the phase of Eq.~(\ref{non parallel vel}) gives respectively the magnitude of
the resulting velocity $\ve{w}$ (because $w=\tanh a$) and specifies the direction $\alpha$ of $\ve{w}$ in the plane $(\ve{v_1},\ve{v_2})$.

In optics the overall reflection coefficient of a Fabry-Perot interferometer can
be obtained by taking into account all virtual paths of light inside the interferometer.\cite{Vigoureux92} The
total probability amplitude for light to be reflected by the system can also be directly obtained (for any
number of interfaces) by using Eq.~(\ref{complaw}). Here
\begin{equation}\label{defR}
R_i= r_i\,e^{i\phi_i}
\end{equation}
is the complex reflection coefficient of an incident wave on the interface $i$, where $r_i$ is the Fresnel coefficient of that
interface and $\phi_i$ is the phase shift corresponding to the propagation of light through the same
homogeneous layer between two successive interfaces. For two interfaces the reflection coefficient of the whole system can be obtained directly by
using the law (\ref{complaw})
\begin{equation}\label{fabry perot}
R=re^{i\phi}=R_2\oplus R_1=\frac{r_1\,e^{i\phi_1}+ r_2\,e^{i\phi_2}}{1+r_2\,e^{-i\phi_2}\,r_1\,e^{i\phi_1}}.
\end{equation}
Again, the modulus and the phase of Eq.~(\ref{fabry perot}) give the overall reflection coefficient and phase of the
reflected wave.

Similarly, we can consider the composition of two non-perfect polarizers $P_1$ and $P_2$.\cite{Lages08} The polarizer $P$ resulting from the combination of polarizers $P_1$ and $P_2$ (in that order) can also be obtained from Eq.~\eqref{complaw}. As explained in Ref. \onlinecite{Lages08}, each polarizer is characterized by
$$
P_i=\tanh\displaystyle\frac{\gamma_i}2\,e^{\,i\alpha_i}
$$
where $\gamma_i$ gives the quality of the polarizer (typically, $\gamma_i=\Gamma_i z$ where $\Gamma_i$ is the differential absorption rate of the polarizer and $z$ the distance traveled by the light wave inside the polarizer; the case of a perfect
polarizer corresponds to $\gamma_i\rightarrow+\infty$), and where $\alpha_i$ gives the orientation of the polarizer axis with respect to an arbitrary reference axis. The polarizer's orientations and the reference axis are coplanar.
The characteristics of the resulting
polarizer $P$ are given by\cite{Lages08}
\begin{equation}\label{polarizers}
P=\tanh\displaystyle\frac{\gamma}{2}\,e^{i\alpha}=P_2\oplus P_1= \frac{\tanh\displaystyle\frac{\gamma_1}{2}\,e^{i\alpha_1}+ \tanh\displaystyle\frac{\gamma_2}{2}\,e^{i\alpha_2}} {1+\tanh\displaystyle\frac{\gamma_1}{2}\,e^{-i\alpha_1}\,\tanh\displaystyle\frac{\gamma_2}{2}\,e^{i\alpha_2}}.
\end{equation}
By using the composition law (\ref{polarizers}) we easily extract the $\gamma$ factor which is the quality of the resulting
polarizer and its direction $\alpha$.

The use of the composition law (\ref{complaw}) is general and can be applied
to any number of coplanar velocities in special relativity, to any number of interfaces for the case of multilayers, and
to any number of successive polarizers. In such cases we have to iterate Eq.~\eqref{complaw} as
relation\cite{Vigoureux92,Lages08,Grossel97}
\begin{equation}\label{complawgen}
A=A_n\oplus(A_{n-1}\oplus\cdots(A_{2}\oplus A_1)).
\end{equation}
The successive iteration of Eq.~(\ref{complaw}) yields the desired result. Equation~(\ref{complawgen})
leads to algorithms which are useful for many problems. It is easy to compute
$A_{2}\oplus A_1$ and then to compose the result with $A_{3}$ and so on.

As explained in Refs.~\onlinecite{Vigoureux91} and \onlinecite{Vigoureux92}, the expression for $A$ in Eq.~(\ref{complawgen}) can be
written down directly by using a complex generalization of the elementary symmetric functions of the variables $A_1$,
$A_2$,$\cdots$, $A_n$ which are extensively used in the theory of polynomials.\cite{Lang65,Waerden66}

Our aim in this paper is to show how Eq.~\eqref{complaw} is related to $2\times2$ matrices and how it provides a
simple way to calculate the four elements of \textit{scattering matrices} ($S$-matrices). We also show how the use of Eq.~(\ref{complaw})
leads naturally to a particular phase, which for the case of the special relativity is related to the Thomas precession.

\section{Matrix representation}
We now explain how the composition law (\ref{complaw}) is related to $2\times2$ matrices.

\begin{figure}[ht!]
\includegraphics[width=8cm]{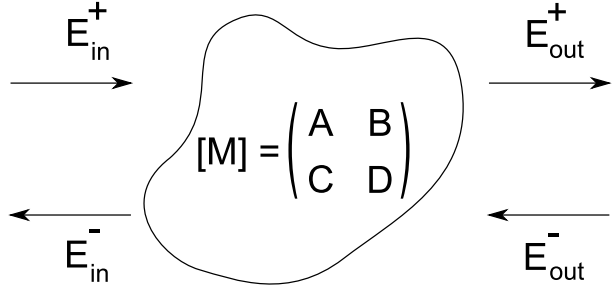}
\caption{Schematic representation of the linear
relations between input and output
quantities.\label{fig1}}
\end{figure}

\subsection{Mathematical definitions}
Consider a physical system (see Fig.~\ref{fig1}) in which two physical quantities (the inputs) $E^+_{\text{in}}$ and
$E^-_{\text{in}}$ are linearly related to two other physical quantities (the outputs), $E_{\text{out}}^+$ and $E_{\text{out}}^-$.
These relations can be written as a $2\times2$ matrix as
\begin{equation}\label{M}
\begin{pmatrix}
E_{\text{in}}^+ \\
E_{\text{in}}^- \\
\end{pmatrix}
=
\begin{pmatrix}
A & B \\
C & D \\
\end{pmatrix}
\begin{pmatrix}
E_{\text{out}}^+ \\
E_{\text{out}}^- \
\end{pmatrix}.
\end{equation}
This general representation is for example used to describe a birefringent system with the help of Jones matrices (see \textit{e.g.} Ref. \onlinecite{Lages08}), or to estimate properties of a multilayer stack with the Abeles matrices.\cite{Abeles50}
It is
possible to define using the $2\times2$ matrix in Eq.~\eqref{M}, hereafter called $[M]$, the four coefficients
$\mathcal{R}^+$, $\mathcal{R}^-$, $\mathcal{T}^+$, and $\mathcal{T}^-$ (this notation is chosen in analogy to the
classical reflection and transmission coefficients of a multilayer device)
\begin{subequations}
\label{R+all}
\begin{align}
\mathcal{R}^+ & = \frac{E_{\text{in}}^-}{E_{\text{in}}^+}\bigg|_{E_{\text{out}}^-=0} =\frac{C}{A},\label{R+}\\
\mathcal{R}^- & = \frac{E_{\text{out}}^+}{E_{\text{out}}^-} \bigg |_{E_{\text{in}}^+=0} =-\frac{B}{A},\label{R-}\\
\mathcal{T}^+ & = \frac{E_{\text{out}}^+}{E_{\text{in}}^+} \bigg |_{E_{\text{out}}^-=0} =\frac{1}{A},\label{T+}\\
\mathcal{T}^- & = \frac{E_{\text{in}}^-}{E_{\text{out}}^-} \bigg |_{E_{\text{in}}^+=0} =\frac{\text{det}[M]}{A}.\label{T-}
\end{align}
\end{subequations}
We introduce the variable $\Theta$ defined by
\begin{equation}
\Theta = \frac{D}{A}\label{eq_theta}.
\end{equation}
If we use Eqs.~(\ref{R+all}) and (\ref{eq_theta}), it is easy to verify that
\begin{equation}\label{stokes} \mathcal{T}^+\,\mathcal{T}^--\mathcal{R}^+\,\mathcal{R}^-=\Theta,
\end{equation}
which constitutes a generalization of the Stokes relation which is well known in the optics of multilayer devices (see \textit{e.g.} Ref. \onlinecite{BornWolf}).
We also introduce the conjugation operation (denoted by the bar):
\begin{subequations}
\label{eqinv12}
\begin{align}
\overline{\mathcal{R}\,}^+ &= -\mathcal{R}^-= \frac{B}{A},\label{eqinv1} \\
\noalign{\noindent and}
\overline{\mathcal{R}\,}^- &= -\mathcal{R}^+.\label{eqinv2}
\end{align}
\end{subequations}
The conjugation operation does not necessarily correspond
to the usual complex conjugation (compare Eqs.~\eqref{R+} and \eqref{eqinv1}). For hermitian matrices the correspondence does hold.

With these definitions the matrix $[M]$ can be written as
\begin{equation}\label{MRT}
[M] = \frac{1}{\mathcal{T}^+}
\begin{pmatrix}
1 & \overline{\mathcal{R}}^+ \\
\mathcal{R}^+ \ & \Theta \\
\end{pmatrix}.
\end{equation}
This form of $[M]$ will be useful in the following derivations.

\subsection{Composition laws of the $\mathcal{R}$ and $\mathcal{T}$ variables}

We now focus on the properties of the four coefficients $\mathcal{R}^+_{21}$, $\mathcal{R}^-_{21}$,
$\mathcal{T}^+_{21}$, and $\mathcal{T}^-_{21}$ of a system characterized by its $[M_{21}]$ matrix, resulting from the composition of two subsystems characterized by the two $[M_1]$ and $[M_2]$ matrices defined by
\begin{equation}\label{M1}
[M_1] = \frac{1}{\mathcal{T}^+_1}
\begin{pmatrix}
1 & \overline{\mathcal{R}}^+_1 \\
\mathcal{R}^+_1 \ & \Theta_1 \\
\end{pmatrix}
\hspace{1cm}\text{and}\hspace{1cm}
[M_2] = \frac{1}{\mathcal{T}^+_2}
\begin{pmatrix}
1 & \overline{\mathcal{R}}^+_2 \\
\mathcal{R}^+_2 \ & \Theta_2 \\
\end{pmatrix}.
\end{equation}
The $[M_{21}]$ matrix is the result of the product of two matrices: $[M_{21}]=[M_2]\,[M_1]$.
Equations~\eqref{eq_theta}--\eqref{eqinv12} allow us to express the composition laws for
$\mathcal{R}^+_{21}$, $\mathcal{T}^+_{21}$, $\mathcal{R}^-_{21}$, and $\mathcal{T}^-_{21}$ as
\begin{subequations}
\label{eq5all}
\begin{align}
\mathcal{R}^+_{21} &= \mathcal{R}_2^+ \oplus\,\mathcal{R}_1^+ =\frac{\mathcal{R}_1^+\,\Theta_2+\mathcal{R}_2^+}{1+\mathcal{R}_1^+\,\overline{\mathcal{R}}_2^+},\label{eq2}\\
\mathcal{R}^-_{21} &= \mathcal{R}_2^- \oplus\,\mathcal{R}_1^- =\frac{\mathcal{R}_1^-+\mathcal{R}_2^-\,\Theta_1}{1+\overline{\mathcal{R}}_1^-\,\mathcal{R}_2^-},\label{eq3}\\
\mathcal{T}^+_{21} &=\mathcal{T}_2^+ \otimes\,\mathcal{T}_1^+ =\frac{\mathcal{T}_1^+\,\mathcal{T}_2^+}{1+\mathcal{R}_1^+\,\overline{\mathcal{R}}_2^+},\label{eq4}\\
\mathcal{T}^-_{21} &=\mathcal{T}_2^- \otimes\,\mathcal{T}_1^- =\frac{\mathcal{T}_1^-\,\mathcal{T}_2^-}{1+\overline{\mathcal{R}}_1^-\,\mathcal{R}_2^-}.\label{eq5}
\end{align}
\end{subequations}
Equation~\eqref{eqinv12} can be used to show that the denominators in Eq.~\eqref{eq5all} are the same,
$1+\mathcal{R}_1^+\,\overline{\mathcal{R}}_2^+ =1+\overline{\mathcal{R}}_1^-\,\mathcal{R}_2^-$.

\subsection{Composition law of the $\Theta$ variables}
Although the $\Theta$ variable has been introduced in an \textit{ad hoc} way in Eq.~(\ref{eq_theta}), it is interesting to find its composition law. We consider two processes characterized by the two variables $\Theta_1$ and $\Theta_2$. If we start from the
generalized Stokes relation (\ref{stokes}) and use Eq.~\eqref{eq5all}, we find
\begin{subequations}
\begin{align}
\Theta_{21} & = \mathcal{T}^+_{21}\,\mathcal{T}^-_{21}-\mathcal{R}^+_{21}\,\mathcal{R}^-_{21}\\
& = \frac{\mathcal{T}_1^+\,\mathcal{T}_2^+\,\mathcal{T}_1^-\,\mathcal{T}_2^--
\left[\mathcal{R}_1^+ \Theta_2 +\mathcal{R}_2^+\right]
\left[\mathcal{R}_1^- +\mathcal{R}_2^- \Theta_1\right]} {\left[1+\mathcal{R}_1^+\,\overline{\mathcal{R}}_2^+\right]^2}.\label{toto}
\end{align}
\end{subequations}
>From Eq.~\eqref{stokes} we know that $\mathcal{T}_1^+ \mathcal{T}_1^-=\Theta_1+\mathcal{R}_1^+\,\mathcal{R}_1^-$ and
$\mathcal{T}_2^+\,\mathcal{T}_2^-=\Theta_2+\mathcal{R}_2^+\,\mathcal{R}_2^-$. Consequently Eq.~\eqref{toto} becomes
\begin{equation}\label{thetavar} \Theta_{21}=\frac{\Theta_1\Theta_2+\overline{\mathcal{R}}_1^+\mathcal{R}_2^+} {1+\mathcal{R}_1^+\overline{\mathcal{R}}_2^+}.
\end{equation}
This expression can be considered as the composition law for $\Theta_1$ and $\Theta_2$.
In Sec.~\ref{sec3} we will give the meaning of $\Theta$ for various physical contexts.

\subsection{$S$-matrix}
By definition, the four coefficients
$\mathcal{R}^+$, $\mathcal{T}^+$, $\mathcal{R}^-$, and $\mathcal{T}^-$ are the four elements of the $S$-matrix associated
with scattering,
\begin{equation}
S= \begin{pmatrix}
\mathcal{R}^+ & \mathcal{T}^- \\
\mathcal{T}^+ & \mathcal{R}^-
\end{pmatrix}.
\end{equation}
Equation~\eqref{eq5all} shows that the composition of two $S$ matrices can be written as
\begin{equation}
S= S_2 \circ S_1 =
\begin{pmatrix}
\mathcal{R}_2^+\oplus\mathcal{R}_1^+ && \mathcal{T}_2^-\otimes\mathcal{T}_1^- \\
\mathcal{T}_2^+\otimes\mathcal{T}_1^+ && \mathcal{R}_2^-\oplus\mathcal{R}_1^-
\end{pmatrix}.
\end{equation}
The use of the composition laws $\oplus$ and $\otimes$  give the elements of the $S$-matrix without resorting to
the usual transfer matrices.

\section{The $\Theta$ phase factor}\label{sec3}

We now consider conservative systems described by the unitary matrix $[U]$.
In this context the $\Theta$ variables are modulus one complex numbers of the form $e^{i\phi}$.
Our aim is to show that the
phases associated with the physical modes $E^+$ and $E^-$ in Eq.~(\ref{M}) can be written as a sum of phases when the two modes are not coupled, plus a phase which is simply expressed with the help of the $\oplus$ law.

\subsection{The composition law in the case of unitary matrices}\label{paragA}
The general expression of a $2\times2$ unitary matrix is
\begin{equation}
[U]=
\begin{pmatrix}
\cos\lambda\,e^{iu} & -\sin\lambda\,e^{iv} \\
\sin\lambda\,e^{-iv} & \cos\lambda\,e^{-iu} \\
\end{pmatrix}
\,e^{i\varphi},
\end{equation}
where $\varphi$, $\lambda$, $u$, and $v$ are real numbers. The overall phase $\varphi$ can be omitted without loss of generality and hereafter we set it equal to zero. As we can see, when the modes are not coupled, that is, when $\lambda=0$, the
evolution matrix reduces to the simple diagonal expression
\begin{equation}\label{matriceu}
[U_{\lambda=0}]=[u]=
\begin{pmatrix}
e^{iu} & 0 \\
0 & e^{-iu} \\
\end{pmatrix}.
\end{equation}
The phase difference between the uncoupled modes $E^+$ and $E^-$ is equal to $2u$. When $\lambda\neq0$, the
evolution of the modes are coupled and the $[U]$ matrix can be factorized as
\begin{subequations}
\label{factorise}
\begin{align}
[U]&= [U_{\lambda=0}][M]=[u][M]\\
&= \begin{pmatrix}
e^{iu} \ & 0 \\
0 & e^{-iu} \\
\end{pmatrix}
\begin{pmatrix}
\cos\lambda & -\sin\lambda \ e^{-i(u-v)} \\
\sin\lambda \ e^{i(u-v)} & \cos\lambda \\
\end{pmatrix}.
\end{align}
\end{subequations}
Such a factorization will help us to estimate the $\mathcal{R}^+$ and $\Theta$ components of the different matrices. By using Eqs.~\eqref{R+}, \eqref{T+} and (\ref{eq_theta}), we find
\begin{equation}
\mathcal{R}_u^+ = 0,\qquad
\mathcal{T}_u^+ = e^{-iu},\qquad
\Theta_u = e^{-2iu},
\end{equation}
so that
\begin{equation}
[u]=\frac{1}{\mathcal{T}_u^+}
\begin{pmatrix}
1&0\\
0&\Theta_u
\end{pmatrix},
\end{equation}
and
\begin{equation}
\mathcal{R}_M^+ = \tan{\lambda}\,e^{i(u-v)},\qquad
\mathcal{T}_M^+ =\frac{1}{\cos\lambda},\qquad
\Theta_M = 1.
\end{equation}
Hence,
\begin{equation}\label{Munitaire}
[M]=\frac{1}{\mathcal{T}_M^+}
\begin{pmatrix}
1 & \cbare{R}_M^+ \\
\mathcal{R}_M^+ \ & 1 \\
\end{pmatrix}.
\end{equation}
Factorizing the free evolution phases as we did in Eq.~\eqref{factorise} will allow us to point
out a new phase expressed with the help of the $\oplus$ composition law. For this purpose consider the $[U_{21}]$ matrix which is
the product of two unitary matrices $[U_1]$ and $[U_2]$,
\begin{equation}
[U_{21}]=[U_2][U_1].
\end{equation}
The factorization of the free evolution phases gives
\begin{subequations}
\label{productUU}
\begin{align}
[U_{21}] &= [u_2][M_2][u_1][M_1] \\
&= [u_2][u_1][u_1]^{-1}[M_2][u_1][M_1]\\
&= [u_2+u_1][M_2(u_1)][M_1]\\
&= [u_2+u_1][M_{21}],
\end{align}
\end{subequations}
where we have defined the diagonal matrix $[u_2+u_1]=[u_2][u_1]$ and noted that
$[M_{21}]=[M_{2}(u_1)][M_1]$ with
\begin{equation}\label{M2u1}
[M_2(u_1)] =[u_1^{-1}][M_2][u_1]=
\begin{pmatrix}
\cos\lambda_2 & -\sin\lambda_2\,e^{-i(2u_1+u_2-v_2)} \\
\sin\lambda_2\,e^{i(2u_1+u_2-v_2)} & \cos\lambda_2\\
\end{pmatrix}.
\end{equation}
If we use the definition (\ref{eq_theta}), we easily find
\begin{equation}
\Theta_{M_1}=\Theta_{M_2}=\Theta_{M_2(u_1)}=1.
\end{equation}
>From the composition law Eq.~(\ref{thetavar}) we obtain
\begin{equation}\label{eq_ph0}
\Theta_{M_{21}}=\frac{\Theta_{M_1}\Theta_{M_2(u_1)}+ \mathcal{R}_{M_2(u_1)}^+\,\cbare{R}_{M_1}^+}{1+\cbare{R}_{M_2(u_1)}^+\,\mathcal{R}_{M_1}^+} =\frac{1+\cbare{R}_{M_1}^+\,\mathcal{R}_{M_2(u_1)}^+}{1+\mathcal{R}_{M_1}^+\,\cbare{R}_{M_2(u_1)}^+},
\end{equation}
or using the composition law definition in Eq.~(\ref{complaw})
\begin{equation}\label{eq_ph} \Theta_{M_{21}}=\frac{\mathcal{R}^+_{M_2(u_1)}\oplus\mathcal{R}^+_{M_1}}{\mathcal{R}^+_{M_1}\oplus\mathcal{R}^+_{M_2(u_1)}}.
\end{equation}

It is interesting to note that $\Theta_{M_{21}}$ comes from the non-commutativity of the composition law
$\oplus$. Although distinct, the two composite quantities $\mathcal{R}^+_{M_1}\oplus \mathcal{R}^+_{M_2(u_1)}$ and
$\mathcal{R}^+_{M_2(u_1)}\oplus \mathcal{R}^+_{M_1}$ have the same modulus, so that $\Theta_{M_{21}}$ is a pure
phase term
\begin{equation}
\Theta_{M_{21}}=e^{-2i\phi}.
\end{equation}
Finally, the whole phase term $\Theta_{U_{21}}=e^{-2i\phi_{21}}$ associated with the $[U_{21}]$ matrix is
\begin{equation} \Theta_{U_{21}}=e^{-2i\phi_{12}}=\Theta_{u_1+u_2}\Theta_{M_{21}}=\Theta_{M_{21}}e^{-2i(u_1+u_2)},
\end{equation}
which gives the phase
\begin{equation}\label{phph}
\phi_{21}=u_1+u_2+\phi.
\end{equation}
The non-commutativity of the $\oplus$ law implies $\Theta_{M_{21}}\neq1$ in Eq.~(\ref{eq_ph}) and is responsible for the additional phase $\phi$ appearing in Eq.~(\ref{phph}).

\subsection{Examples of the physical meaning of the $\Theta$ variable}
In the following we illustrate the meaning of the phase term $\Theta$ by three examples
from different fields of physics.

\subsubsection{Special relativity}

We first choose the composition of two non-parallel velocities $\ve{v_1}$ and $\ve{v_2}$. In this case
the four elements $A$, $B$, $C$, and $D$ of the matrix (\ref{M}) are respectively $\cosh(a_i/2)$,
$\sinh(a_i/2)\, e^{-i\alpha_i}$, $\sinh(a_i/2)\,e^{i\alpha_i}$, and
$\cosh a_i/2$ where $a_i$ and $v_i=\tanh a_i$ are respectively the rapidity and the velocity of the reference frame $i$ for a given observer.
Equations \eqref{R+} and \eqref{eqinv1} then give
$V_i=\mathcal{R}^+_i=\displaystyle\tanh{(a_i/2)}\,e^{i\alpha_i}$ and
$\overline{V}_i=\cbare{R}^+_i=\displaystyle\tanh{(a_i/2)}\,e^{-i\alpha_i}$. Here the phase $\alpha_i$ gives the
orientation of $\ve{v_i}$ with respect to an arbitrary axis belonging to the plane defined by the vectors $\ve{v_1}$ and $\ve{v_2}$ in the observer reference frame. Because $\Theta_i=1$, Eqs.~\eqref{thetavar}, \eqref{eq_ph0}, and \eqref{eq_ph} give
\begin{equation}\label{thomas} \Theta_{21}=\frac{1+\cbare{R}_1^+\,\mathcal{R}_2^+}{1+\mathcal{R}_1^+\,\cbare{R}_2^+}= \frac{1+\overline{V}_1\,V_2}{1+V_1\,\overline{V}_2}= \displaystyle\frac{1+\tanh\displaystyle\frac{a_1}{2}\,e^{-i\alpha_1}\,\tanh\displaystyle\frac{a_2}{2}\,e^{i\alpha_2}} {1+\tanh\displaystyle\frac{a_1}{2}\,e^{i\alpha_1}\,\tanh\displaystyle\frac{a_2}{2}\,e^{-i\alpha_2}}.
\end{equation}
This expression is a pure phase term and can be written as
\begin{equation}\label{thomas^2}
\Theta_{21}=e^{-2i\phi},
\end{equation}
where $2\phi$ is the Wigner's angle associated with the Thomas precession. Note that from Eq.~\eqref{thomas} we
directly obtain the value of the Wigner angle. The real part of $\Theta_{21}$ gives immediately the known result\cite{Vigoureux01,Wigner39}
\begin{equation}\label{wigner} \cos{2\phi}=\displaystyle\frac{(1+\tanh\displaystyle\frac{a_1}{2}\tanh\displaystyle\frac{a_2}{2}\,\cos{\alpha})^2- (\tanh\displaystyle\frac{a_1}{2}\tanh\displaystyle\frac{a_2}{2}\,\sin{\alpha})^2} {(1+\tanh\displaystyle\frac{a_1}{2}\tanh\displaystyle\frac{a_2}{2}\,\cos{\alpha})^2+ (\tanh\displaystyle\frac{a_1}{2}\tanh\displaystyle\frac{a_2}{2}\,\sin{\alpha})^2},
\end{equation}
where $\alpha=\alpha_2-\alpha_1$.

As was shown at the end of Sec.~\ref{paragA}, the Wigner angle comes from the
non-commutativity of the composition law which mimics the non-commutativity of the Lorentz boosts. The $\oplus$ law can be easily used for the composition of any number of
co-planar velocities. For example, for three referential frames, we obtain by iterating Eq.~\eqref{complaw}
\begin{equation}\label{3V}
W=V_3\oplus(V_2\oplus V_1)=\displaystyle\frac{V_1+V_2+V_3+V_1\bare{V}_2\,V_3} {1+\bare{V}_1\,V_2+\bare{V}_1\,V_3+\bare{V}_2\,V_3}.
\end{equation}

\subsubsection{The optics of stratified media}

This example is from the optics of stratified media. If $r_i$ and $t_i$ are the Fresnel reflection
and transmission coefficients of the interface $i$ ($i=1,2$), and $\phi_i$ is the phase shift associated with the
propagation of light, the four elements $A$, $B$, $C$, and $D$ of the matrix (\ref{M}) are respectively
$1/t_i$, $(r_i/t_i)\,e^{-i\phi_i}$, $(r_i/t_i)\,e^{i\phi_i}$, and
$1/t_i$. Equation~\eqref{R+} gives $\mathcal{R}_i^+ = r_i\,e^{i\phi_i}$, so that Eq.~\eqref{eq2} gives the
overall reflection coefficient\cite{BornWolf} of the two interfaces (\ref{fabry perot}).
In this case a phase
term\cite{Vigoureux98} also appears which is strictly similar to the Wigner angle in special relativity.
Its origin comes also from the non-commutativity of the $\oplus$ law which is related to the
non-invariance of the problem when the two interfaces are exchanged.

\subsubsection{Light wave polarization}

As our last example, we consider two non-perfect polarizers.
As explained in Sec.~\ref{somex}, the quality of the polarizer resulting from using successively two polarizers $P_1$ and $P_2$ is given\cite{Lages08}
 by Eq.~\eqref{complaw}. It is interesting in this case to calculate the value of
$\Theta_{21}$. As was explained, $\Theta_{21}$ expresses the non-commutativity of the two quantities which
are \textit{composed}. In special relativity finding two different results when calculating the resulting velocity of
$v_1$ composed with $v_2$ and of $v_2$ composed with $v_1$ might have been surprising. It is
not the case with polarizers. It is well known that the final polarization of a light wave going through the polarizer $P_1$ and then through the polarizer $P_2$ is not the same as the final polarization of the light wave going first through polarizer $P_2$ and then through polarizer $P_1$. We
consider explicitly the non-commutativity of polarizers. From Eqs.~\eqref{polarizers} and \eqref{thetavar} we obtain
\begin{equation}\label{non com pol} e^{-2i\,\Omega}=\displaystyle\frac{1+\tanh\displaystyle\frac{\gamma_1}{2}\,e^{i\alpha_1}\, \tanh\displaystyle\frac{\gamma_2}{2}\,e^{-i\alpha_2}} {1+\tanh\displaystyle\frac{\gamma_1}{2}\,e^{-i\alpha_1}\,\tanh\displaystyle\frac{\gamma_2}{2}\,e^{i\alpha_2}}.
\end{equation}
In Eq.~\eqref{non com pol} $2\Omega$ is the angle between the polarization of light $\ve{E}_{12}$ when going through the two
polarizers in the order $P_1$ and then $P_2$ and that of light $\ve{E}_{21}$ when going through the polarizers in the
order $P_2$ and $P_1$. For two perfect polarizers we expect to find
$2 \Omega=\alpha_2-\alpha_1=\alpha$. To verify this result, replace $\tanh\displaystyle\frac{\gamma_1}{2}$ and
$\tanh\displaystyle\frac{\gamma_2}{2}$ by unity for perfect polarizers, and then the corresponding Eq.~\eqref{non com pol} for polarizers
gives the expected result
\begin{equation}
\cos{2\,\Omega}=\cos\alpha.
\end{equation}

\section{Discussion}
>From Eq.~\eqref{Munitaire}, we observe that there are redundancies of information in $2\times2$ unitary matrices. All
information is contained in the first (or the second) column. Because of this redundancy, it is easy to understand why
the use of the composition law (\ref{complaw}) is easier and more rapid than using matrix methods such as transfer
matrices. Moreover, as shown in Ref.~\onlinecite{Grossel94}, calculations converge more rapidly
when the composition law is used. This rapid convergence comes from the fact that the denominator of the composition law is a normalization
factor.
Another useful aspect of the composition law is that it can be easily iterated
\begin{equation}\label{complawR}
\mathcal{R}^+_{n,\ldots,1}= \mathcal{R}^+_n\oplus (\mathcal{R}^+_{n-1}\oplus\cdots(\mathcal{R}^+_{2}\oplus \mathcal{R}^+_{1})).
\end{equation}
As mentioned, this property leads to efficient algorithms for many kinds of problems. Also, Eq.~(\ref{complawR}) is so simple that its analytic value can be directly given
without any matrix calculations. Equation~(\ref{complawR}) is a complex generalization of the
elementary symmetric functions of the mathematical theory of polynomials:\cite{Vigoureux93} the numerator of
$\mathcal{R}^+_{n,\cdots,1}$ is constituted by all the possible odd ordered products of the different $\mathcal{R}^+_i$ factors, such that in each
product, the $\mathcal{R}^+$ and $\overline{\mathcal{R}}^+$ factors appear alternatively, the first factor always being
$\mathcal{R}^+$. The denominator of $\mathcal{R}^+_{n,\cdots,1}$is constituted by all the possible even ordered products of $\mathcal{R}^+_i$, such that in each product, the $\overline{\mathcal{R}}^+$ and $\mathcal{R}^+$ factors appear
alternatively, the first always being $\overline{\mathcal{R}}^+$. If we limit ourselves to two iterations, the value of
$\mathcal{R}^+_{3,\cdots,1}$ is directly given by
\begin{subequations}
\label{R4}
\begin{align}
\mathcal{R}^+_{3,\cdots,1} &= \mathcal{R}^+_3\oplus(\mathcal{R}^+_2\oplus\mathcal{R}^+_1)\\
&= \frac{\mathcal{R}^+_1+\mathcal{R}^+_2+\mathcal{R}^+_3+\mathcal{R}^+_1\cbare{R}^+_2\mathcal{R}^+_3} {1+\cbare{R}^+_1\mathcal{R}^+_2+\cbare{R}^+_1\mathcal{R}^+_3+\cbare{R}^+_2\mathcal{R}^+_3}.
\end{align}
\end{subequations}
Such a result can be useful for the case of $S$-matrices because the generalization of Eq.~(\ref{R4}) allows us to write the $S$-matrix simply as
\begin{equation}
S=
\begin{pmatrix}
\mathcal{R}_n ^+\oplus\cdots(\mathcal{R}_2^+\oplus\mathcal{R}_1^+)\quad\quad & \mathcal{T}_n^-\otimes\cdots(\mathcal{T}_2^-\otimes\mathcal{T}_1^-)\\
& \\
\mathcal{T}_n^+\otimes\cdots(\mathcal{T}_2^+\otimes\mathcal{T}_1^+)\quad\quad & \mathcal{R}_n^-\oplus\cdots(\mathcal{R}_2^-\oplus\mathcal{R}_1^-)
\end{pmatrix}.
\end{equation}
It is well known that a number of physical processes are more adequately described by
$S$-matrices than by $T$-matrices (transfer matrices). Unfortunately, whereas $T$-matrices must be successively multiplied together, $[M_{n,\cdots,1}]=[M_n]\,[M_{n-1}]\cdots [M_2]\,[M_1]$, such is not the case with $S$-matrices. The
composition law is consequently useful for $S$-matrices because our results show how to directly calculate
the four elements of the overall $S$-matrix by iterating Eq.~\eqref{eq5all}.

\section{Conclusion}
The composition law of velocities in special relativity appears to be the natural way to
add velocities that are subject to the condition $|v| < c$. Its generalization in the complex
plane leads to simple calculations of bounded quantities which would be otherwise difficult to calculate.
We have shown how, for example, the Wigner angle in special relativity, the overall reflection
coefficient of any multilayer, and the effect of any number of polarizers can be directly obtained from this general composition law. Also, we have shown that the generalization of the Einstein's composition law provides a natural way to compose scattering matrices.


\begin{thebibliography}{99}
\bibitem{footnote}The $\gamma$'s are proportional to the absorption rate of their respective polarizer; the case of a perfect
polarizer corresponds to $\gamma\rightarrow\infty$.
\end{thebibliography}
\end{document}